\documentstyle[12pt,amssymb,amsmath]{article}
\makeatletter
\renewcommand{\@biblabel}[1]{}
\makeatother

\begin{document}

\title{ {\bf The quasiclassical limit of the symmetry constraint of
the KP hierarchy and the dispersionless KP hierarchy with
self-consistent sources}  }

\author{ {\bf  Ting Xiao
    \hspace{1cm} Yunbo Zeng\dag } \\
    {\small {\it
    Department of Mathematical Sciences, Tsinghua University,
    Beijing 100084, China}} \\
    {\small {\it \dag
     Email: yzeng@math.tsinghua.edu.cn}}}

\date{}
\maketitle
\renewcommand{\theequation}{\arabic{section}.\arabic{equation}}

\begin{abstract}
For the first time we show that the quasiclassical limit of the
symmetry constraint of the KP hierarchy leads to the generalized
Zakharov reduction of the dispersionless KP (dKP) hierarchy which
has been proved to be result of symmetry constraint of the dKP
hierarchy recently. By either regarding the constrained dKP
hierarchy as its stationary case or taking the dispersionless
limit of the KP hierarchy with self-consistent sources directly,
we construct a new integrable dispersionless hierarchy, i.e., the
dKP hierarchy with self-consistent sources and find its associated
conservation equations (or equations of Hamilton-Jacobi type).
Some solutions of the dKP equation with self-consistent sources
are also obtained by hodograph transformations.
\end{abstract}

\hskip\parindent

{\bf{Keywords}}: quasiclassical limit; constrained KP hierarchy;
(dispersionless) KP hierarchy with self-consistent sources;
conservation equation (equation of Hamilton-Jacobi type); Zakharov
reduction; hodograph transformation

\section{Introduction}
\setcounter{equation}{0} \hskip\parindent The dispersionless
integrable hierarchies provide us an interesting type of nonlinear
integrable models which have important applications from complex
analysis to topological field theory (see [1-12] and the
references therein). In
\cite{Lebedev1979,Zakharov1980,Kodama1988,Kodama1989,Takasaki1995},
a standard procedure of dispersionless limit of integrable
dispersionfull hierarchies is proposed. In this procedure,
dispersionless hierarchies arise as the quasiclassical limit of
the original dispersionfull Lax equations performed by replacing
operators by phase space functions, commutators by Poisson
brackets and the role of Lax pair equations by conservation
equations (or equations of Hamilton-Jacobi type ). A
$\bar{\partial}$ scheme of dispersionless hierarchies has been
proposed by Konopelchenko et al (see \cite{Konopelchenko2001,
Bogdanov2004} and the references therein). Recently, from this
point of view, some important reductions of dispersionless
hierarchies are shown to be nothing but symmetry constraints
\cite{Bogdanov2004}. Also several methods for solving
dispersionless hierarchies have been formulated such as twistorial
method \cite{Takasaki1995,Martinez2003} and hodograph
transformation \cite{Kodama1988,Kodama1989}. In
\cite{Kodama1988,Kodama1989}, from the conservation equations of
the dispersionless KP equation, Kodama and Gibbons found exact
solutions of it and its reductions by hodograph transformations
and obtained general hodograph equations for hydrodynamical type
equations.\\

The soliton equations with self-consistent sources (SESCS) are
another type of important integrable models in many fields of
physics, such as hydrodynamics, state physics, plasma physics, etc
(see [14-18]). For example, the KP equation with self-consistent
sources (KPESCS) describes the interaction of a long wave with a
short-wave packet propagating on the x,y plane at an angle to each
other (see \cite{Mel'nikov1987} and the references therein). In
general, the constrained integrable hierarchy may be viewed as
stationary system of the corresponding hierarchy with
self-consistent sources and the Lax representation for the latter
can be obtained naturally from that for the former [16-18]. In
this sense, the soliton hierarchies with self-consistent sources
may be viewed as integrable generalizations
of the original soliton hierarchies. \\

In contrast with the dispersionfull integrable hierarchies with
self-consistent sources, the dispersionless integrable hierarchies
with self-consistent sources have not been studied before. In this
paper, we first investigate the quasiclassical limit of the
symmetry constraint of the KP hierarchy which leads to the
generalized Zakharov reduction of the dKP hierarchy. The latter
has recently been proved to be symmetry constraint of the dKP
hierarchy using the $\bar{\partial}$ method \cite{Bogdanov2004}.
So, we find the relation that the symmetry constraint for the Lax
function of the dispersionless hierarchy can be obtained by the
quasiclassical limit of the symmetry constraint for the Lax
operator of the corresponding dispersionfull hierarchy. By either
regarding the constrained dKP hierarchy as its stationary case or
taking the dispersionless limit of the KP hierarchy with
self-consistent sources directly, we construct a new
dispersionless hierarchy, i.e., the dKP hierarchy with
self-consistent sources. This hierarchy is also integrable since
we can find its associated conservation equations (or equations of
Hamilton-Jacobi type). So in this sense, the dKP hierarchy with
self-consistent sources may be viewed as an integrable
generalization of the dKP hierarchy. Compared with the case of a
nonlocal term appearing in the $t$-part of the Lax pair for the KP
equation with self-consistent sources \cite{Xiao20041}, the
$t$-part of the conservation equations for the dKP equation with
self-consistent sources possesses rational terms with poles. From
the obtained conservation equations, we can solve the dKP
hierarchy with self-consistent sources by hodograph
transformations.\\

The paper will be organized as follows. In section 2, we remind
some definitions and results about the KP hierarchy with
self-consistent sources. In section 3, we take the quasiclassical
limit to the symmetry constraint of the KP hierarchy to obtain the
symmetry constraint of the dKP hierarchy. The dKP hierarchy with
self-consistent sources and its associated conservation equations
are obtained. Section 4 is devoted to solving the dKP hierarchy
with self-consistent sources by hodograph transformations and some
solutions for the dKP equation with self-consistent sources are
presented.

\section{The KP hierarchy with self-consistent sources}
\setcounter{equation}{0} \hskip\parindent We first review some
definitions and results about the KP hierarchy with
self-consistent sources in the framework of Sato theory [17-20].
Given a pseudo-differential operator (PDO) of the form
\begin{equation}
\label{1}
    L=\partial+u_1(t)\partial^{-1}+u_2(t)\partial^{-2}+...,
\end{equation}
where $\partial=\partial_x$,
$\partial\partial^{-1}=\partial^{-1}\partial=1$,
$t=(t_1=x,t_2,\cdots)$, the KP hierarchy is defined as
\begin{equation}
\label{2}
    \partial_{t_n}L=[B_n,L],
\end{equation}
where $B_n=(L^n)_{\geq 0}$ is the differential part of $L^n$. The
Lax euqation (\ref{2}) is equivalent to the existence of the Baker
function $\psi$ such that
\begin{subequations}
\label{3}
\begin{equation}
\label{3.1}
    L\psi = \lambda\psi,
\end{equation}
\begin{equation}
\label{3.2}
    \partial_{t_n}\psi = B_n\psi,
\end{equation}
\end{subequations}
and $\psi$ also satisfies
\begin{equation}
\label{3.3}
    \partial_\lambda\psi=M\psi,
\end{equation}
where M is the Orlov operator of the KP hierarchy. The adjoint
Baker function $\psi^*$ satisfies
\begin{subequations}
\label{4}
\begin{equation}
\label{4.1}
    L^*\psi^* = \lambda\psi^*,
\end{equation}
\begin{equation}
\label{4.2}
    \partial_n\psi^* = -B_n^*\psi^*,
\end{equation}
\begin{equation}
\label{4.3}
    \partial_\lambda\psi^* = -M^*\psi^*.
\end{equation}
\end{subequations}

Making a constraint of the PDO $L$ (\ref{1}) as \cite{Oevel1993}
\begin{equation}
\label{5}
    L^n=(L^n)_{\geq 0}+\sum_1^Nq_i(t)\partial^{-1}r_i(t),\ \ \
    n\in \mathbb{N}
\end{equation}
where $q_i(t)$ and $r_i(t)$ satisfying
\begin{equation}
\label{6}
    q_{i,t_m} = B_mq_i,\ \ r_{i,t_m} = -B_m^*r_i,\ \ B_m =
    [(L^n)^{\frac{m}{n}}]_{\geq 0},\ \ i=1,...,N, \forall
    m\in\mathbb{N},
\end{equation}
we will get the $n$-constrained KP hierarchy as
\begin{subequations}
\label{7}
\begin{equation}
\label{71}
     (L^n)_{t_m}=[B_m,L^n],
\end{equation}
\begin{equation}
\label{b012}
    q_{i,t_m}=B_mq_i,
\end{equation}
\begin{equation}
\label{b013}
    r_{i,t_m}=-B_m^*r_i,\ \ \ i=1,...,N.
\end{equation}
\end{subequations}
If we add the term $(B_m)_{t_n}$ to the right side of equation
(\ref{71}), the KP hierarchy with self-consistent sources will be
obtained as \cite{Xiao20041}
\begin{subequations}
\label{8}
\begin{equation}
\label{81}
     (B_m)_{t_n}-(L^n)_{t_m}+[B_m,L^n]=0,
\end{equation}
\begin{equation}
\label{82}
    q_{i,t_m}=B_mq_i,
\end{equation}
\begin{equation}
\label{83}
    r_{i,t_m}=-B_m^*r_i,\ \ \ i=1,...,N.
\end{equation}
\end{subequations}
As many cases in (1+1)-dimension (see \cite{Zeng2000} and the
references therein), the $n$-constrained KP hierarchy may be
considered as the stationary one of the KP hierarchy with
self-consistent sources if "$t_n$" is viewed as the evolution
variable and the Lax representation for the KP hierarchy with
self-consistent sources can be obtained naturally from that for
the constrained KP hierarchy \cite{Xiao20041}. The most important
equation in the hierarchy is the KP equation with self-consistent
sources obtained when $n=3$, $m=2$ in (\ref{8}),
\cite{Mel'nikov1987}
\begin{subequations}
\label{9}
\begin{equation}
\label{91}
     [u_{1,t}-3u_1u_{1,x}-\frac{1}{4}u_{1,xxx}+\sum_{i=1}^N(q_ir_i)_x]_x-\frac{3}{4}u_{1,yy} = 0,
\end{equation}
\begin{equation}
\label{92}
    q_{i,y} = q_{i,xx}+2u_1q_i,
\end{equation}
\begin{equation}
\label{93}
    r_{i,y} = -r_{i,xx}-2u_1r_i, \ \ \ i=1,...,N
\end{equation}
\end{subequations}
where $t=t_3, y=t_2$. With (\ref{92}) and (\ref{93}), (\ref{91})
will be obtained by the compatibility of the following linear
equations (Lax pair)\cite{Xiao20041}
\begin{subequations}
\begin{equation}
\psi_y = \psi_{xx}+2u_1\psi,
\end{equation}
\begin{equation}
\psi_t =
\psi_{xxx}+3u_1\psi_x+\frac{3}{2}(u_{1,x}+(\partial^{-1}u_{1,y}))\psi+\sum_{i=1}^Nq_i\partial^{-1}(r_i\psi).
\end{equation}
\end{subequations}

\section{Dispersionless limit}
\setcounter{equation}{0} \hskip\parindent Following the standard
procedure of dispersionless limit introduced in
\cite{Lebedev1979,Zakharov1980,Kodama1988,Kodama1989,Takasaki1995},
we will get the dispersionless counterpart of (\ref{8}) which can
be regarded as the dispersionless KP hierarchy with
self-consistent sources. Simply taking $T_n=\epsilon t_n$ and
thinking of $u_n(\frac{T}{\epsilon})=U_n(T)+O(\epsilon)$ as
$\epsilon\rightarrow 0$, $L$ in (\ref{1}) changes into
\begin{equation}
\label{10}
    L_\epsilon=\epsilon\partial+\sum_{i=1}^\infty
    u_i(\frac{T}{\epsilon})(\epsilon\partial)^{-i}=\epsilon\partial+\sum_{i=1}^\infty
    (U_i(T)+O(\epsilon))(\epsilon\partial)^{-i},\ \ \ \
    \partial=\partial_X,\ \ \ X=\epsilon x.
\end{equation}
The constraint (\ref{5}) now changes into
\begin{equation}
\label{11}
    L^n_\epsilon=B_{\epsilon n}+\sum_{i=1}^Nq_i(\frac{T}{\epsilon})(\epsilon\partial)^{-1}r_i(\frac{T}{\epsilon}),\ \ B_{\epsilon n}=(L^n_\epsilon)_{\geq
    0},
\end{equation}
where $q_i(\frac{T}{\epsilon})$ and $r_i(\frac{T}{\epsilon})$
satisfy
\begin{equation}
\label{12}
    \epsilon [q_i(\frac{T}{\epsilon})]_{T_m} = B_{\epsilon
    m}q_i(\frac{T}{\epsilon}),\ \ \epsilon [r_i(\frac{T}{\epsilon})]_{T_m} = -B^*_{\epsilon
    m}r_i(\frac{T}{\epsilon}),\ \ B_{\epsilon
    m}=[(L^n_\epsilon)^{\frac{m}{n}}]_{\geq 0},\ \ i=1,...,N,
\end{equation}
and the counterpart of (\ref{7}) is
\begin{subequations}
\label{13}
\begin{equation}
\label{131}
     \epsilon(L^n_\epsilon)_{T_m}=[B_{\epsilon m},L^n_\epsilon],
\end{equation}
\begin{equation}
\label{132}
    \epsilon [q_i(\frac{T}{\epsilon})]_{T_m} = B_{\epsilon
    m}q_i(\frac{T}{\epsilon}),
\end{equation}
\begin{equation}
\label{133}
    \epsilon [r_i(\frac{T}{\epsilon})]_{T_m} = -B^*_{\epsilon
    m}r_i(\frac{T}{\epsilon}),\ \ i=1,...,N.
\end{equation}
\end{subequations}
It is proved in \cite{Takasaki1995} that
\begin{equation}
\label{14}
    {\mathcal{L}} = \sigma^{\epsilon}(L_\epsilon) = p+\sum_{i=1}^\infty
    U_i(T)p^{-i}.
\end{equation}
is a solution of the dKP hierarchy, i.e., satisfies
\begin{equation}
\label{15}
    \partial_{T_n}{\mathcal{L}} = \{{\mathcal{B}}_n, {\mathcal{L}}\},
\end{equation}
where $\sigma^\epsilon$ denotes the principal symbol
\cite{Takasaki1995}, the Poisson bracket is defined as
\begin{equation}
\{A(p,x), B(p,x)\} = \frac{\partial A}{\partial p} \frac{\partial
B}{\partial x}-\frac{\partial A}{\partial x}\frac{\partial
B}{\partial p},
\end{equation}
and ${\mathcal{B}}_n=({\mathcal{L}})^n_{\geq 0}$ now refers to
powers
of $p$.\\
The dKP hierarchy can be also written in the zero-curvature form
\begin{equation}
\label{16}
    \frac{\partial {\mathcal{B}}_n}{\partial T_m}-\frac{\partial {\mathcal{B}}_m}{\partial
    T_n}+\{{\mathcal{B}}_n,{\mathcal{B}}_m\}=0.
\end{equation}
In \cite{Takasaki1995}, from (\ref{3}) and (\ref{3.3}) (with $L$,
$B_n$, $M$ and $\partial_n$ replaced by $L_\epsilon$, $B_{\epsilon
n}$, $M_\epsilon$ and $\epsilon \partial_{T_n}$ respectively ), it
has been proved that $\psi(\frac{T}{\epsilon})$ has the following
WKB asymptotic expansion as $\epsilon\rightarrow 0$
\begin{equation}
\label{17}
    \psi(\frac{T}{\epsilon})=exp[\frac{1}{\epsilon}S(T,\lambda)+O(1)],\
    \ \ \epsilon\rightarrow 0.
\end{equation}
In order to take the quasiclassical limit of the constraint
(\ref{11}), we also need to find the asymptotic form of the
adjoint Baker function. Similarly like the proof in
\cite{Takasaki1995}, from (\ref{4}), we can find
$\psi^*(\frac{T}{\epsilon})$ has the following WKB asymptotic
expansion
\begin{equation}
\label{18}
    \psi^*(\frac{T}{\epsilon})=exp[-\frac{1}{\epsilon}S(T,\lambda)+O(1)],\
    \ \ \epsilon\rightarrow 0.
\end{equation}
From (\ref{3.2}) and (\ref{17}), we obtain a hierarchy of
conservation equations for the momentum function $p=\frac{\partial
S}{\partial X}$,
\begin{equation}
\label{19}
    \frac{\partial p}{\partial T_n}=\frac{\partial {\mathcal{B}}_n(p)}{\partial
    X},
\end{equation}
the compatibility of which, i.e., $\frac{\partial^2 p}{\partial
T_n \partial T_m}=\frac{\partial^2 p}{\partial T_m \partial T_n}$
implies the dKP hierarchy (\ref{16}). \\
Regarding
\begin{subequations}
\label{20}
\begin{equation}
\label{201}
     q_i(\frac{T}{\epsilon})=\psi(\frac{T}{\epsilon},\lambda=\lambda_i) \sim
     exp[\frac{S(T,\lambda_i)}{\epsilon}+\alpha_{i1}+O(\epsilon)],\
     \ \epsilon\rightarrow 0,
\end{equation}
\begin{equation}
\label{202}
    r_i(\frac{T}{\epsilon})=\psi^*(\frac{T}{\epsilon},\lambda=\lambda_i) \sim
     exp[-\frac{S(T,\lambda_i)}{\epsilon}+\alpha_{i2}+O(\epsilon)],\
     \ \epsilon\rightarrow 0,\ \ i=1,...,N,
\end{equation}
\end{subequations}
we will find that
\begin{equation}
\nonumber
\begin{array}{ll}
    &q_i(\frac{T}{\epsilon})(\epsilon
    \partial)^{-1}r_i(\frac{T}{\epsilon})\\=&e^{\alpha_{i1}+\alpha_{i2}}[(\epsilon\partial)^{-1}+(\frac{\partial S(T,\lambda_i)}{\partial X}+O(\epsilon))(\epsilon
    \partial)^{-2}+((\frac{\partial S(T,\lambda_i)}{\partial X})^2+O(\epsilon))(\epsilon\partial)^{-3}+\cdots],\ \ \epsilon\rightarrow 0.
\end{array}
\end{equation}
Taking the principal symbol of both sides of (\ref{11}), we have
\begin{eqnarray}
\label{21}
{\mathcal{L}}^n&=&{\mathcal{B}}_n+\sum_{i=1}^Ne^{\alpha_{i1}+\alpha_{i2}}[p^{-1}+S_X(T,\lambda_i)p^{-2}+S_X^2(T,\lambda_i)p^{-3}+\cdots]\nonumber\\
&=&{\mathcal{B}}_n+\sum_{i=1}^N\frac{v_i}{p-p_i}
\end{eqnarray}
where ${\mathcal{B}}_n=({\mathcal{L}}^n)_{\geq 0}$ and
\begin{equation}
\label{22} v_i=e^{\alpha_{i1}+\alpha_{i2}},\ \
p_i=S_X(T,\lambda_i).
\end{equation}
The constraint (\ref{21}) is well known and is often called
Zakharov reduction when $n=1$
\cite{Zakharov1980,Krichever1992,Aoyama1996,Bogdanov2004}. In
\cite{Bogdanov2004}, using the $\bar{\partial}$ method, the
authors demonstrated that (\ref{21}) is a result of symmetry
constraint of the dKP hierarchy. Here we have shown that
(\ref{21}) can be obtained by the dispersionless limit of
(\ref{5}), i.e., the dispersionless limit of symmetry constraint
for the Lax operator of the dispersionfull hierarchy leads to the
symmetry constraint for the Lax function of the dispersionless
hierarchy. From (\ref{12}), (\ref{20}), (\ref{22}) and by a
tedious computation, we obtain the following equations of
hydrodynamical type
\begin{subequations}
\label{23}
\begin{equation}
\label{231}
     p_{i,T_k}=[{\mathcal{B}}_k(p)|_{p=p_i}]_X,
\end{equation}
\begin{equation}
\label{232}
    v_{i,T_k}=[v_i(\frac{\partial}{\partial p}{\mathcal{B}}_k(p))|_{p=p_i}]_X,\ \
    i=1,\cdots,N.
\end{equation}
\end{subequations}
Taking the dispersionless limit of (\ref{13}), we will get the
constrained dKP hierarchy reduced by (\ref{21})
\begin{subequations}
\label{24}
\begin{equation}
\label{241}
     ({\mathcal{L}}^n)_{T_k} = \{{\mathcal{B}}_k,{\mathcal{L}}^n\},
\end{equation}
\begin{equation}
\label{242}
     p_{i,T_k}=[{\mathcal{B}}_k(p)|_{p=p_i}]_X,
\end{equation}
\begin{equation}
\label{243}
    v_{i,T_k}=[v_i(\frac{\partial}{\partial p}{\mathcal{B}}_k(p))|_{p=p_i}]_X,\ \
    i=1,...,N.
\end{equation}
\end{subequations}
Adding the term $({\mathcal{B}}_k)_{T_n}$ to the right hand side
of (\ref{241}), or taking the dispersionless limit of (\ref{8})
directly, we will obtain the dKP hierarchy with self-consistent
sources as
\begin{subequations}
\label{25}
\begin{equation}
\label{251}
     ({\mathcal{B}}_k)_{T_n}-({\mathcal{L}}^n)_{T_k}+ \{{\mathcal{B}}_k,{\mathcal{L}}^n\}=0,
\end{equation}
\begin{equation}
\label{252}
     p_{i,T_k}=[{\mathcal{B}}_k(p)|_{p=p_i}]_X,
\end{equation}
\begin{equation}
\label{253}
    v_{i,T_k}=[v_i(\frac{\partial}{\partial p}{\mathcal{B}}_k(p))|_{p=p_i}]_X,\ \
    i=1,...,N.
\end{equation}
\end{subequations}
If "$T_n$" is viewed as the evolution variable, (\ref{24}) may be
regarded as the stationary system of (\ref{25}) like the
dispersionfull case. It is not difficult to prove that under
(\ref{252}) and (\ref{253}), the equation (\ref{251}) will be
obtained by the compatibility of the following conservation
equations
\begin{subequations}
\label{26}
\begin{equation}
\label{261}
     p_{T_k}=({\mathcal{B}}_k(p))_X,
\end{equation}
\begin{equation}
\label{262}
     p_{T_n}=({\mathcal{L}}^n(p))_X=[{\mathcal{B}}_n(p)+\sum_{i=1}^N\frac{v_i}{p-p_i}]_X.
\end{equation}
\end{subequations}
For example, when $n=3$, $k=2$,
\begin{equation}
\nonumber
     {\mathcal{L}}^3=p^3+3U_1p+3U_2+\sum_{i=1}^N\frac{v_i}{p-p_i},\
     \ \ {\mathcal{B}}_2 = p^2+2U_1,
\end{equation}
(\ref{25}) becomes the dKP equation with self-consistent sources
($U_2$ is eliminated by $U_{2,X}=\frac{1}{2}U_{1,Y}$ and $T=T_3$,
$Y=T_2$)
\begin{subequations}
\label{27}
\begin{equation}
\label{271}
     (U_{1,T}-3U_1U_{1,X}+\sum_{i=1}^Nv_{i,X})_X=\frac{3}{4}U_{1,YY},
\end{equation}
\begin{equation}
\label{272}
     p_{i,Y}=(p_i^2+2U_1)_X,
\end{equation}
\begin{equation}
\label{273}
    v_{i,Y}=2(v_ip_i)_X,\ \ i=1,...,N.
\end{equation}
\end{subequations}
Under (\ref{272}) and (\ref{273}), the compatibility of the
following conservation equations give rise to (\ref{271})
\begin{subequations}
\label{28}
\begin{equation}
\label{281}
     p_Y=(p^2+2U_1)_X=2pp_X+2U_{1,X},
\end{equation}
\begin{equation}
\label{282}
     p_T=(p^3+3U_1p+3U_2+\sum_{i=1}^N\frac{v_i}{p-p_i})_X=3p^2p_X+3(U_1p)_X+3U_{2,X}+\sum_{i=1}^N\frac{v_{i,X}}{p-p_i}-\sum_{i=1}^N\frac{v_i(p_X-p_{i,X})}{(p-p_i)^2}.
\end{equation}
\end{subequations}
where $U_{2,X}=\frac{1}{2}U_{1,Y}$.

\section{Hodograph solutions} \setcounter{equation}{0} \hskip\parindent
Motivated by the dKP case, we would like to solve (\ref{25}) by
hodograph transformation provided the conservation equations
(\ref{26}). Following \cite{Kodama1988}, one can consider the
$M$-reductions of (\ref{26}) so that the momentum function $p$ and
$v_i$, $p_i$,$i=1,...,N$ depend only on a set of functions
$W=$($W_1$,...,$W_M$) with $W_1=U_1$ and ($W_1$,...,$W_M$) satisfy
commuting flows
\begin{equation}
\label{29} \frac{\partial W}{\partial T_n} = A_n(W)\frac{\partial
W}{\partial X},\ \ n\geq 2
\end{equation}
where the $M\times M$ matrices $A_n$ are functions of
$(W_1,...,W_M)$ only. In the following, we shall take the dKP
equation with self-consistent sources (\ref{27}) for example and
show its solutions in the cases of $M=1$ and $M=2$.\\
1. $M=1$\\
In this case, $p=p(U_1)$, $v_i=v_i(U_1)$, $p_i=p_i(U_1)$ and
(\ref{29}) becomes
\begin{equation}
\label{30}
     U_{1,Y}=A(U_1)U_{1,X},\ \ \ U_{1,T}=B(U_1)U_{1,X}.
\end{equation}
From (\ref{272}), (\ref{273}) and (\ref{28}), we will get the
following relations respectively
\begin{subequations}
\label{31}
\begin{equation}
\label{311}
     \frac{d p_i}{d U_1}(\frac{A(U_1)}{2}-p_i) = 1,
\end{equation}
\begin{equation}
\label{312}
 \frac{d v_i}{d U_1}(\frac{A(U_1)}{2}-p_i) = v_i\frac{d p_i}{d U_1},
\end{equation}
\begin{equation}
\label{313}
     \frac{d p}{d U_1}(\frac{A(U_1)}{2}-p) = 1,
\end{equation}
\begin{equation}
\label{314}
     \frac{d p}{d U_1}B(U_1)=3p^2\frac{d p}{d U_1}+3p+3U_1\frac{d p}{d U_1}+3\frac{d U_2}{d U_1}+\sum_{i=1}^N\frac{\frac{d v_i}{d U_1}}{p-p_i}-\sum_{i=1}^Nv_i\frac{(\frac{d p}{d U_1}-\frac{d p_i
     }{d U_1})}{(p-p_i)^2},
\end{equation}
\end{subequations}
which implies
\begin{equation}
\label{32}
     B=3U_1+\frac{3}{4}A^2-\sum_{i=1}^N\frac{d v_i}{d
     U_1},\ \ \ A=2\frac{d U_2}{d U_1}.
\end{equation}
It is easy to verify that with (\ref{31}) and (\ref{32}),
(\ref{30}) are compatible. Making the hodograph transformations
with the change of variables $(X,Y,T)\rightarrow (U_1,Y,T)$ and
$X=X(U_1,Y,T)$, we will get the following hodograph equations for
$X$,
\begin{equation}
\label{33}
     \frac{\partial X}{\partial Y}=-A,\ \ \ \frac{\partial X}{\partial
     T}=-B=-3U_1-\frac{3}{4}A^2+\sum_{i=1}^N\frac{d v_i}{d
     U_1},
\end{equation}
which can be easily integrated as follows
\begin{equation}
\label{34}
     X+A(U_1)Y+(3U_1+\frac{3}{4}A(U_1)^2-\sum_{i=1}^N\frac{d v_i}{d U_1})T =
     F(U_1),
\end{equation}
where $F(U_1)$ is an arbitrary function of $U_1$.\\
If we choose $A(U_1)=C_0=$ const, $F(U_1)=\beta U_1$, from
(\ref{34}), (\ref{311}) and (\ref{312}), we get an implicit
solution as
\begin{subequations}
\label{35}
\begin{equation}
\label{351}
      X+C_0Y+[3U_1+\frac{3}{4}C_0^2-\sum_{i=1}^Nc_i(\frac{C_0^2}{4}-2(U_1+d_i))^{-\frac{3}{2}}]T =\beta U_1,
\end{equation}
\begin{equation}
\label{352} v_i=c_i[\frac{C_0^2}{4}-2(U_1+d_i)]^{-\frac{1}{2}},
\end{equation}
\begin{equation}
\label{353}
     p_i=\frac{C_0}{2}\pm\sqrt{\frac{C_0^2}{4}-2(U_1+d_i)},\ \
     i=1,...,N,
\end{equation}
\end{subequations}
where $c_i$,$d_i$,$i=1,...,N$ are constants. If
$d_1=d_2=\cdots=d_N$ and $\sum_{i=1}^Nc_i=0$, (\ref{351})
degenerates to the solution of dKP equation \cite{Kodama1988}.\\
If we assume $v_1=v_2=\cdots=v_N$ and $\sum_{i=1}^N\frac{d v_i}{d
U_1}=3U_1$, i.e., $\frac{d v_i}{d U_1}=\frac{3}{N}U_1$, and taking
$F(U_1)=0$, by a direct computation, we will obtain an explicit
solution of the dKP equation with self-consistent sources
(\ref{27}) as follows
\begin{equation*}
\nonumber U_1=\frac{4Y^2-6TX\pm4Y\sqrt{Y^2-3TX}}{225T^2},
\end{equation*}
\begin{equation*}
\nonumber v_i=\frac{3}{2N}U_1^2,
\end{equation*}
\begin{equation*}
\nonumber
 p_i=2\sqrt{2U_1},\ \ i=1,...,N.
\end{equation*}
\\
2. $M=2$\\
In this case we denote $W_1=U_1$, $W_2=W$, then $v_i=V_i(U_1,W)$,
$p_i=p_i(U_1,W)$ and $p=p(U_1,W)$ with the commuting flows
\begin{equation}
\label{36}
\begin{pmatrix} U_1\\W
\end{pmatrix}_Y=A\begin{pmatrix} U_1\\W
\end{pmatrix}_X,\ \ \ \ \ \ \ \begin{pmatrix} U_1\\W
\end{pmatrix}_T=B\begin{pmatrix} U_1\\W
\end{pmatrix}_X,
\end{equation}
where $A=(A_{ij})$ and $B=B_{ij}$ are $2\times 2$ matrix functions
of $U_1$ and $W$. Requiring $U_{1,x}$ and $W_x$ are independent,
($\ref{272}$), ($\ref{273}$) and ($\ref{28}$) give rise to the
following relations respectively
\begin{subequations}
\label{37}
\begin{equation}
\label{371}
     (\frac{\partial p_i}{\partial U_1}, \frac{\partial p_i}{\partial W})A=(2+2p_i\frac{\partial p_i}{\partial U_1}, 2p_i\frac{\partial p_i}{\partial W}),
\end{equation}
\begin{equation}
\label{372}
 (\frac{\partial v_i}{\partial U_1}, \frac{\partial v_i}{\partial W})A=(2\frac{\partial (v_ip_i)}{\partial U_1}, 2\frac{\partial (v_ip_i)}{\partial W}),
\end{equation}
\begin{equation}
\label{373}
    (\frac{\partial p}{\partial U_1},\frac{\partial p}{\partial W})A=2p(\frac{\partial p}{\partial U_1}, \frac{\partial p}{\partial W})+(2,0),
\end{equation}
\begin{equation}
\label{374}
\begin{array}{lll}
 (\frac{\partial p}{\partial U_1},\frac{\partial p}{\partial W})B&=&3p^2(\frac{\partial p}{\partial U_1}, \frac{\partial p}{\partial W})+3(\frac{\partial (U_1p)}{\partial U_1},\frac{\partial (U_1p)}{\partial W})+3(\frac{\partial U_2}{\partial U_1},\frac{\partial U_2}{\partial W})\\
 &&+\sum_{i=1}^N\frac{1}{p-p_i}(\frac{\partial v_i}{\partial U_1},\frac{\partial v_i}{\partial W})-\sum_{i=1}^N\frac{v_i}{(p-p_i)^2}(\frac{\partial p}{\partial U_1}-\frac{\partial p_i}{\partial U_1},\frac{\partial p}{\partial W}-\frac{\partial p_i}{\partial
 W}),
\end{array}
\end{equation}
\end{subequations}
which implies $A(U_1,W)$ and $B(U_1,W)$ must satisfy
\begin{equation}
\label{38} B=\frac{3}{4}A^2+3U_1I-\sum_{i=1}^N\frac{\partial
v_i}{\partial U_1}I-\begin{pmatrix} 0 & \sum_{i=1}^N\frac{\partial
v_i}{\partial W}\\ \frac{A_{11}}{A_{12}}\sum_{i=1}^N\frac{\partial
v_i}{\partial W}&
\frac{A_{22}-A_{11}}{A_{12}}\sum_{i=1}^N\frac{\partial
v_i}{\partial W} \end{pmatrix}
\end{equation}
with $A_{11}=2\frac{\partial U_2}{\partial U_1}$ and
$A_{12}=2\frac{\partial U_2}{\partial W}$.\\
For simplicity we assume $\frac{\partial v_i}{\partial W}=0$,
$i=1,...,N$. Then
\begin{equation}
\label{39}
 B=\frac{3}{4}A^2+3U_1I-\sum_{i=1}^N\frac{\partial
v_i}{\partial
U_1}I=\frac{3}{4}(trA)A+3(U_1-\frac{1}{4}detA-\frac{1}{3}\sum_{i=1}^N\frac{\partial
v_i}{\partial U_1})I,
\end{equation}
where the formula $A^2=(trA)A-(detA)I$ is used.\\
With (\ref{39}), the compatibility for (\ref{36}) requires $A$ to
satisfy
\begin{equation}
\label{40}
\begin{pmatrix} \frac{1}{4}\frac{\partial detA}{\partial
W}\\\frac{\partial(U_1-\frac{1}{4}detA-\frac{1}{3}\sum_{i=1}^N\frac{\partial
v_i}{\partial U_1})}{\partial U_1}
\end{pmatrix}=A\begin{pmatrix} \frac{1}{4}\frac{\partial trA}{\partial
W}\\-\frac{1}{4}\frac{\partial trA}{\partial U_1}\end{pmatrix}.
\end{equation}
Using the hodograph transformation changing the independent
variables ($X,Y,T$) to ($U_1,W,T$) with $X=X(U_1,W,T)$ and
$Y=Y(U_1,W,T)$, we get
\begin{equation}
\label{41}
\begin{pmatrix} -X_W\\X_{U_1}
\end{pmatrix}=A\begin{pmatrix} Y_W\\-Y_{U_1}
\end{pmatrix},\ \ \ \ \ \ \ \begin{pmatrix} \frac{\partial (X,Y)}{\partial
(W,T)}\\-\frac{\partial (X,Y)}{\partial (U_1,T)}
\end{pmatrix}=B\begin{pmatrix} Y_W\\-Y_{U_1}\end{pmatrix},
\end{equation}
where $\frac{\partial (X,Y)}{\partial (W,T)}=X_WY_T-X_TY_W$. It is
not difficult to see that (\ref{41}) has solutions in the form
\begin{subequations}
\label{42}
\begin{equation}
\label{421}
     X+3(U_1-\frac{1}{4}detA-\frac{1}{3}\sum_{i=1}^N\frac{\partial v_i}{\partial
     U_1})T=F(U_1,W),
\end{equation}
\begin{equation}
\label{422} Y+\frac{3}{4}(trA)T=G(U_1,W),
 \end{equation}
\end{subequations}
where we have required that $Y_{U_1}$ and $Y_W$ are independent,
while $F$ and $G$ satisfy the linear equations
\begin{equation}
\label{43}
\begin{pmatrix} -F_W\\F_{U_1}
\end{pmatrix}=A\begin{pmatrix} G_W\\-G_{U_1}
\end{pmatrix}.
\end{equation}
An example of solution is given by
\begin{equation}
\label{44} A=\begin{pmatrix} W&U_1\\
4&W
\end{pmatrix},
\end{equation}
and
\begin{equation}
\label{45} v_i=c_iU_1,\ \  p_i=\frac{W}{2}, \ \ i=1,\cdots,N,
\end{equation}
with $c_i$,$i=1,\cdots,N$ are constants.\\
(\ref{42}) now becomes
\begin{subequations}
\label{46}
\begin{equation}
\label{461}
     X+3[U_1-\frac{1}{4}(W^2-4U_1)-\frac{1}{3}\sum_{i=1}^Nc_i]T=F(U_1,W),
\end{equation}
\begin{equation}
\label{462} Y+\frac{3}{4}2WT=G(U_1,W).
 \end{equation}
\end{subequations}
From (\ref{43}) and $F_{WU_1}=F_{U_1W}$, $G$ must satisfy
$$2G_{U_1}+U_1G_{U_1U_1}-4G_{WW}=0.$$
Taking $G=-W$ and from (\ref{43}), $F=\frac{1}{2}W^2-4U_1$ and we
obtain a solution of (\ref{27}) as follows
\begin{subequations}
\label{47}
\begin{equation}
\label{471}
U_1=\frac{Y^2}{2(3T+2)^2}+\frac{(\sum_{i=1}^Nc_i)T-X}{2(3T+2)},
\end{equation}
\begin{equation}
\label{472}
v_i=c_iU_1=c_i[\frac{Y^2}{2(3T+2)^2}+\frac{(\sum_{i=1}^Nc_i)T-X}{2(3T+2)}],
\end{equation}
\begin{equation}
\label{473} p_i=\frac{W}{2}=-\frac{Y}{3T+2}.\ \ i=1,...,N.
\end{equation}
\end{subequations}
(\ref{47}) is a global solution for $T>-\frac{2}{3}$ and
(\ref{471}) degenerates to the solution of the dKP equation
\cite{Kodama1988} when $\sum_{i=1}^Nc_i=0$.

\section*{Acknowledgment}\hskip\parindent
This work was supported by the Chinese Basic Research Project
"Nonlinear Science".

\hskip\parindent

\end{document}